\documentclass[12pt]{iopart}

\usepackage{graphicx}
\usepackage{dcolumn}
\usepackage{bm}
\usepackage{epsfig}
\usepackage{cite}
\newcommand{\be}{\begin{eqnarray}}
\newcommand{\ee}{\end{eqnarray}}

\begin{document}

\title[Phase Diagrams of Three-Component Attractive Fermions]{Phase Diagrams of Three-Component 
Attractive Ultracold Fermions in One-Dimension}

\author{ C.C.N. Kuhn and  A. Foerster }

\address{Instituto de Fisica da UFRGS, Av. Bento Gon\c{c}alves, 9500, Porto Alegre, 91501-970,Brazil}

\eads{\mailto{carlos.kuhn@ufrgs.br}, \mailto{angela@if.ufrgs.br}}

\begin{abstract}
We investigate trions, paired states and quantum phase transitions in one-dimensional $SU(3)$ attractive fermions 
in external fields by means of the Bethe ansatz formalism.
Analytical  results for the ground state energy, critical fields and complete phase diagrams are 
obtained for the weak coupling regime. Higher order corrections for these physical quantities are 
presented in the strong attractive regime.
Numerical solutions of the dressed energy equations allow us to examine how the different phase boundaries 
modify by varying the inter-component coupling throughout the whole attractive regime.
The pure trionic phase existing in the strong coupling regime reduces smoothly by decreasing 
this coupling until the weak limit is reached. 
In this weak regime, a pure BCS-like paired phase can be sustained under certain nonlinear Zeeman splittings.
\end{abstract}

\pacs{ 02.30.Ik, 03.75.Ss, 03.75.Hh, 64.70.Tg}

\submitto{\NJP}

\maketitle

\section{Introduction}
Recent experiments on ultracold atomic systems confined to one dimension (1D) \cite{1D-F,weiss,g2,Haller} 
have attracted renewed interest in Bethe ansatz integrable models of interacting bosons and multi-component fermions. 
The most recent experimental breakthrough  is  the realization of a 1D spin-imbalanced Fermi gas of $^6$Li atoms under the degenerate
temperature \cite{Hulet}.
This study demonstrates how ultracold atomic gases in 1D may be used to create non-trivial new phases of matter, and also 
paves the way for direct observation and further  study of the 
Fulde-Ferrell-Larkin-Ovchinnikov (FFLO) states \cite{FFLO,MF-FFLO}. 

Three-component fermions exhibit a rich scenario, revealing more exotic phases \cite{Rapp,Demler,Lecheminant2,Wilczek,Paananen,Rapp2,Thai,Silva}.
Notably, strongly attractive three-component ultracold atomic fermions can form three-body bound states called {\it trions}.
Consequently, a phase transition  between Bardeen-Cooper-Schrieffer (BCS) like pairing superfluid and trionic states 
is expected to occur in the strong attractive regime \cite{Lecheminant2,Rapp2,xiwen,Hui,XiwenYupeng,BAXK,xw,pf,inaba}.
So far, most of the theoretical analysis has focused on the attractive strong coupling limit. 
A pertinent discussion in this context is what happens at intermediate and weak attractive coupling 
regimes 

In this paper we consider one-dimensional three-component ultracold fermions with $\delta$-function 
interaction in  external magnetic fields.
From a mathematical point of view, this model was solved long ago by Sutherland \cite{Sutherland} and 
Takahashi \cite{Takahashi} through Bethe ansatz techniques.
Recently, integrable models of three-component interacting fermions \cite{xiwen,Hui,Errea} have received renewed interest 
in connection with ultracold atomic gases.
Advanced experimental techniques newly developed allow to explore three-component Fermi gas with different phases of 
trions, dimers and free atoms \cite{Li,Li2,shuta,knoop}. Remarkably, direct observation of a trimer state consisting of 
fermionic $^6$Li atoms in the three energetically lowest substates has been just reported in \cite{sciencejochim}.
This opens up an opportunity to experimentally study such novel quantum phases of trions and pairs in 1D 
three-component Fermi gases, providing a physical ground and stimulus for the investigation of their theoretical 
aspects.

Our aim here is to expand on the theoretic knowledge of 1D integrable model of three-component fermions by undertaking a 
detailed analysis of how the different phases modify as the inter-component interaction decreases, ranging from a 
strong to a weak regime. 
We obtain analytical expressions for the critical fields and construct the full phase diagrams in the weak coupling 
regime by solving the Bethe ansatz equations.
We extend previous work on this model \cite{xiwen,BAXK} to derive higher order corrections for these physical quantities 
in the strong coupling regime.
Numerical solutions of the dressed energy equations show that the pure trionic phase existing in the strong coupling 
regime reduces as the coupling decreases and it disappears in the presence of external fields when the 
weak regime is approached. 
In contrast to the two-component interacting fermions \cite{jingsong}, nonlinear Zeeman splitting may sustain a
BCS-like paired phase in the three-component attractive fermions for weak coupling regime. 

\section{The model}  

We consider a $\delta$-function (contact potential) interacting system of $N$ fermions with equal mass $m$, which can occupy three 
possible hyperfine levels ($|1\rangle$, $|2\rangle$ and $|3\rangle$) with particle number $N^1$, $N^2$ and $N^3$, respectively. 
They are constrained to a line of length $L$ with periodic boundary conditions. The Hamiltonian reads \cite{Sutherland} 
\begin{eqnarray}
{H}&=&-\frac{\hbar ^2}{2m}\sum_{i = 1}^{N}\frac{\partial
^2}{\partial x_i^2}+\,g_{\rm 1D} \sum_{1\leq i<j\leq N} \delta
(x_i-x_j) + E_Z
 \label{Ham}
\end{eqnarray}
The first and second terms correspond to the kinetic energy and $\delta$-interaction potential, respectively.
The last term denotes the Zeeman energy $E_Z=\sum^3_{i=1}N^i\epsilon^{i}_Z(\mu^i_B,B)$, with the Zeeman 
energy levels $\epsilon^{i}_Z$  determined by the magnetic moments $\mu_B^{i}$ and the magnetic field $B$. 
For later convenience, the Zeeman energy term can also be written as $E_Z=-H_1(N^1-N^2)-H_2(N^2-N^3) +N\bar{\epsilon}$, 
where the unequally spaced Zeeman splitting in three hyperfine levels can be specified by two independent 
parameters $H_1 = \bar{\epsilon} - \epsilon^{1}_Z(\mu_B^{1},B)$ and $H_2 = \epsilon^{3}_Z(\mu_B^{3},B)- \bar{\epsilon}$, 
with $\bar{\epsilon}=\sum_{\sigma=1}^3\epsilon^{\sigma}_Z(\mu_B^{i},B)/3$ the average Zeeman energy.

The spin-independent contact interaction $g_{\rm 1D}$ remains between fermions with different hyperfine 
states and preserves the spins in each hyperfine states, i.e., the number of fermions in each spin states are conserved. 
Although these conditions seem rather restrictive, it is possible to tune scattering lengths between atoms in 
different low sublevels to form nearly $SU(3)$  degeneracy Fermi gases via broad Feshbach resonances.\cite{1D-F,Grimm,Li,Li2}.
Consequently the model still captures the essential physics relevant in the discussion of multiple phases in three-component ultracold Fermi gases. 
The inter-component coupling  $g_{\rm 1D} =-{\hbar ^2 c}/{m}$ with interaction strength $c=-{2}/{a_{\rm 1D}}$ is determined by the
effective 1D scattering length $a_{\rm 1D}$ \cite{1d-a}. It is attractive for $g_{\rm 1D}<0$ and repulsive for $g_{\rm 1D}>0$.
Here we will focus in the attractive case. 
For simplicity, we choose the dimensionless units of $\hbar = 2m = 1$ and use the dimensionless coupling constant $\gamma=c/n$ with linear density $n ={N}/{L}$.

The Hamiltonian (\ref{Ham}) exhibits spin $SU(3)$ symmetry and
was solved long ago by means of the nested Bethe ansatz \cite{Sutherland,Takahashi}. In this approach its spin content 
was incorporated via the symmetry of the wavefunction. The energy eigenspectrum is given in terms of the quasimomenta
$\left\{k_i\right\}$  of the fermions through

\begin{equation}
E=\sum_{j=1}^Nk_j^2,
\label{eigs}
\end{equation}
satisfying the following set of coupled Bethe ansatz equations (BAE) \cite{Sutherland,Takahashi}
\begin{eqnarray}
exp(\mathrm{i}k_jL)=\prod^{M_1}_{\ell = 1} \frac{k_j-\Lambda_\ell+\mathrm{i}\, c/2}{k_j-\Lambda_\ell-\mathrm{i}\, c/2},\nonumber \\ 
\prod^N_{\ell = 1}\frac{\Lambda_{\alpha}-k_{\ell}+\mathrm{i}\, c/2}{\Lambda_{\alpha}-k_{\ell}-\mathrm{i}\, c/2}= - {\prod^{M_1}_{ \beta = 1} } \frac{\Lambda_{\alpha}-\Lambda_{\beta} +\mathrm{i}\, c}{\Lambda_{\alpha}-\Lambda_{\beta} -\mathrm{i}\, c} {\prod^{M_2}_{\ell = 1} }\frac{\Lambda_{\alpha}-\lambda_{\ell} -\mathrm{i}\, c/2}{\Lambda_{\alpha}-\lambda_{\ell} +\mathrm{i}\, c/2},\nonumber \\ 
 {\prod^{M_1}_{ \ell = 1} }\frac{\lambda_{\mu}-\Lambda_{\ell} +\mathrm{i}\, c/2}{\lambda_{\mu}-\Lambda_{\ell} -\mathrm{i}\, c/2}=-{\prod^{M_2}_{\ell = 1} }\frac{\lambda_{\mu}-\lambda_{\ell} +\mathrm{i}\, c}{\lambda_{\mu}-\lambda_{\ell} -\mathrm{i}\, c},
\label{BE}
\end{eqnarray}
written also in terms of the rapidities for the internal hyperfine spin degrees 
of freedom $\Lambda_{\alpha}$ and $\lambda_{\mu}$.
Above $j=1,\ldots, N$, $\alpha = 1,\ldots, M_1$, $\mu=1,\ldots,M_2$, with quantum numbers $M_1=N_2+2N_3$ and $M_2=N_3$.
For the irreducible representation $\left[3^{N_3}2^{N_2}1^{N_1}\right]$, a three-column Young tableau 
encodes the numbers of unpaired fermions ($N_1=N^1-N^2$), bound pairs ($N_2=N^2-N^3$) and trions ($N_3=N^3$).

\section{Ground states} 

For attractive interaction, the BAE allow charge bound states and spin strings. In particular, the $SU(3)$ symmetry carries 
two kinds of charge bound states: trions and pairs.
In principle, different numbers of unpaired fermions, pairs and trions can be chosen to populate the ground state 
by carefully tuning $H_1$ and $H_2$.

In the strong coupling regime $L|c|\gg 1$, the imaginary parts of the bound states become equal-spaced, i.e., a 
trionic state has the form $ \left\{k_j=\Lambda_j \pm \mathrm{i}|c|, \lambda_j \right\} $ and for the bound 
pair $ \left\{ k_{r}=\Lambda_{r} \pm \mathrm{i}|c|/2 \right\}$.  
Substituting these root patterns into the BAE (\ref{BE}), we find their real parts, from which the ground 
state energy in the strongly attractive regime can be obtained \cite{xiwen}
\begin{eqnarray}
\frac{E}{L} &\approx& \frac{\pi^2n_1^3}{3}\left(1+\frac{8n_2+4n_3}{|c|}+\frac{12}{c^2}(2n_2+n_3)^2\right) -\frac{n_2c^2}{2}\nonumber\\
& & +\frac{\pi^2n_2^3}{6}\left(1+\frac{12n_1+6n_2+16n_3}{3|c|} + \frac{1}{3c^2}(6n_1+3n_2+8n_3)^2\right)-2n_3c^2 \nonumber\\
& & +\frac{\pi^2n_3^3}{9}\left(1+\frac{12n_1+32n_2+18n_3}{9|c|}+\frac{1}{27c^2}(6n_1+16n_2+9n_3)^2\right).
\label{E}
\end{eqnarray}
Here $n_a=N_a/L$ ($a=1,2,3$) is the density for unpaired fermions, pairs and trions, respectively.
This state can be considered as a mixture of unpaired fermions, pairs and trionic fermions, behaving basically 
like particles with
different statistical signatures \cite{GBLB}.
For strong attractive interaction, trions are stable compared to the BCS-like pairing and unpaired states. 
From (\ref{E}) we can obtain the binding energy for a trion, given by $\varepsilon_t=\hbar^2c^2/m$ and the 
pair binding energy, which is $\varepsilon_b=\hbar^2c^2/4m$.

In the weak coupling regime $L|c|\ll 1$ the imaginary parts $\mathrm{i}y$ of the charge bound states are the roots of Hermite polynomials $H_k$ of degree $k$.
Specifically, $H_k(\sqrt{\frac{L}{2|c|}}y)=0$, with $k=2,3$ for a bound pair and a trion, respectively \cite{su4}. 
The real parts of the quasimomenta deviate smoothly from the values evaluated for the $c=0$ case.
With this root configuration, the ground state energy in the weak attractive regime can be obtained 
\begin{eqnarray}
 \frac{E}{L} &\approx& \frac{\pi^2}{3}(n_1^3+2n_2^3+3n_3^3) \nonumber\\
& & +\pi^2(n_1(n_2+n_3)(n_1+n_2+n_3)+ 2n_2n_3(n_2+n_3)) \nonumber\\
& & -2|c|(n_1n_2+2n_1n_3+4n_2n_3+n_2^2+3n_3^2).
\label{Eweak}
\end{eqnarray}

The ground state energy (\ref{Eweak}) is dominated by the kinetic energy of composite particles and 
unpaired fermions and has an interaction energy consisting of density density interaction between  charge 
bound states and between charge bound states and unpaired fermions.  
From equation (\ref{Eweak}) we can obtain the binding energy for a trion, given by $\varepsilon_t=3\hbar^2|c|/mL$ 
and the pair binding energy, which is $\varepsilon_b=\hbar^2|c|/mL$. 
For weak attractive interaction, the trionic state is unstable against thermal and spin fluctuations. This  
becomes apparent in the weak coupling phase diagrams presented in Fig. \ref{fig:E}(d), Fig. \ref{n1weak}(c) 
and Fig. \ref{n2weak}(c) below.
 
\section{Dressed energy formalism}  

In the thermodynamic limit, i.e. $L,N\to \infty$ with $N/L$  finite, the grand partition function 
$Z=tr(\mathrm{e}^{-\cal{H}/T})=\mathrm{e}^{-G/T}$ is given in terms of the Gibbs free energy 
$G = E + E_{\rm Z} - \mu N - TS$, written in terms of the Zeeman energy $E_{\rm Z}$, chemical 
potential $\mu$, temperature $T$ and entropy $S$ \cite{Takahashi-B,Schlot,BGOS,xiwen}.
The Gibbs free energy can be expressed in terms of the densities of particles and holes for unpaired 
fermions, bound pairs and trions, as well as spin degrees of freedom, which are determined from 
the BAE (\ref{BE}).
Thus the equilibrium state is established by minimizing the Gibbs free energy with 
respect to these densities. This procedure leads to a set of coupled nonlinear integral equations, 
from which the dressed energy equations are obtained in 
the limit $T\to 0$ (see \cite{BGOS,xiwen} for details)
\begin{eqnarray}
\label{eps}
\epsilon^{(3)}(\lambda)&=&3\lambda^2-2c^2-3\mu-a_2*{\epsilon^{(1)}}(\lambda) \nonumber\\
& &-\left[a_1+a_3\right]*{\epsilon^{ (2)}}(\lambda)-\left[a_2+a_4\right]*{\epsilon^{ (3)}}(\lambda) \nonumber\\
\epsilon^{(2)}(\Lambda)&=&2\Lambda^2-2\mu-\frac{c^2}{2}-H_2-a_1*{\epsilon^{ (1)}}(\Lambda)  \nonumber\\
& &-a_2*{\epsilon^{2}}(\Lambda) -\left[a_1+a_3\right]*{\epsilon^{(3)}}(\Lambda) \label{TBA-F}\\
\epsilon^{ (1)}(k)&=&k^2-\mu-H_1-a_1*{\epsilon^{ (2)}}(k) -a_2*{\epsilon^{(3)}}(k). \nonumber
\end{eqnarray}
Here $\epsilon^{(a)}, a=1,2,3$ are the dressed energies for unpaired fermions, bound pairs and trions, respectively and 
$a_j(x)=\frac{1}{2\pi}\frac{j|c|}{(jc/2)^2+x^2}$.
The symbol ``$*$''  denotes the convolution $a_j*{\epsilon^{(a)}}(x)=\int_{-Q_a}^{+Q_a}a_j(x-y){\epsilon^{(a)}}(y)dy$
with the integration boundaries $Q_{a}$ given by $\epsilon^{(a)}(\pm Q_{a})=0$. 
The Gibbs free energy per unit length at zero temperature can be written in terms of the dressed energies 
as $ G = \sum_{a=1}^3 \frac{a}{2\pi} \int_{-Q_a}^{+Q_a}{\epsilon^{(a)}}(x)dx$.

The dressed energy equations (\ref{TBA-F}) can be analytically solved just in some special limits.
In particular, they were solved in \cite{xiwen} for strongly attractive interaction 
through a lengthy iteration method.
Here we numerically solve these equations to determine the full phase diagram of the model for any value of the coupling.
This allows to examine how the different phase boundaries deform by varying the coupling from strong to weak regime. 
The numerical solution is also employed to confirm the analytical expressions for the physical quantities and the resulting phase diagrams of the 
model in the weak coupling limit (see \cite{BAXK} for a similar discussion in the strong regime).

\section{Full phase diagrams} 
Basically, there are two possible Bethe ansatz schemes to construct the phase diagram of the system. 
One possibility is to handle with the dressed energy equations (\ref{TBA-F}).
This approach was discussed for the strong attractive regime in \cite{xiwen}, where expressions for 
the fields in terms of the densities were obtained up to order of $1/|c|$.
Alternatively, one can handle directly with its discrete version (equations (\ref{eigs}) and (\ref{BE})) 
by solving the BAE. We adopt this second strategy here.
In order to obtain the explicit forms for the fields in terms of the polarizations we consider the energy for arbitrary 
population imbalances  

\begin{eqnarray}
E/L=\mu n+G/L +n_1H_1+n_2 H_2,
\label{energy_pot}
\end{eqnarray}
which coincides with the ground state energy (\ref{eigs}) obtained by solving the BAE (\ref{BE}).
Then the fields $H_1$ and $H_2$ are determined through the relations
\begin{eqnarray}
H_1 = \frac{\partial E/L}{\partial n_1}\,,\, H_2=\frac{\partial
E/L}{\partial n_2}\label{termod}
\end{eqnarray}
together with the constraint
\begin{equation}
n = n_1 + 2 n_2 + 3 n_3.
\label {constraint}
\end{equation}

In the strong coupling regime, using the ground state energy (\ref{E}) we find
\begin{eqnarray}
\hspace{-2.5cm}
H_1&=&\pi^2n_1^2\left( 1 - \frac{4n_1}{9|c|} + \frac{8n_2}{|c|} + \frac{4n_3}{|c|} + \frac{12}{c^2}(2n_2+n_3)^2 - \frac{8n_1}{3c^2}(2n_2+n_3)\right)  \nonumber \\
\hspace{-2.5cm}
&-&\frac{\pi^2n^2_3}{9}\left(1+\frac{4n_1}{3|c|} + \frac{32n_2}{9|c|} + \frac{4n_3}{3|c|} + \frac{(6n_1+16n_2+9n_3)^2}{27c^2} - \frac{2n_3(6n_1+16n_2+9n_3)}{9c^2}\right) \nonumber \\ 
\hspace{-2.5cm}
&+&\frac{10\pi^2n^3_2}{27|c|}\left(1+\frac{6n_1+3n_2+8n_3}{|c|}\right) + \frac{2c^2}{3}. \nonumber \\
\hspace{-2.5cm}
H_2&=&\frac{\pi^2n_2^2}{2}\left(1+\frac{4n_1}{|c|} + \frac{40n_2}{27|c|} + \frac{16n_3}{3|c|} + \frac{(6n_1+3n_2+8n_3)^2}{3c^2} - \frac{14n_2(6n_1+3n_2+8n_3)}{27c^2}\right)  \nonumber \\
\hspace{-2.5cm}
&-& \frac{2\pi^2n_3^2}{9}\left(1+\frac{4n_1}{3|c|} + \frac{32n_2}{9|c|} + \frac{8n_3}{9|c|} + \frac{(6n_1+16n_2+9n_3)^2}{27c^2}- \frac{10n_3(6n_1+16n_2+9n_3)}{27c^2}\right) \nonumber \\
\hspace{-2.5cm}
&+& \frac{16\pi^2n_1^3}{9|c|}\left(1+\frac{6(2n_2+n_3)}{|c|}\right) + \frac{5c^2}{6}.
\label{H1-H2}
\end{eqnarray}
These equations provide higher order corrections to those derived in \cite{xiwen} using the dressed energy equations.
For determining the full phase boundaries, we also need the energy-field transfer relation between the paired 
and unpaired phases $H_1-H_2/2$, which can be extracted from the underlying two-component system with $SU(2)$ symmetry 
\cite{GBLB,jingsong}.
These equations determine the full phase diagram and the critical fields activated by the fields $H_{1}$ and $H_{2}$.

%
%
%
\begin{figure}[t]
{{\includegraphics [width=0.5\linewidth]{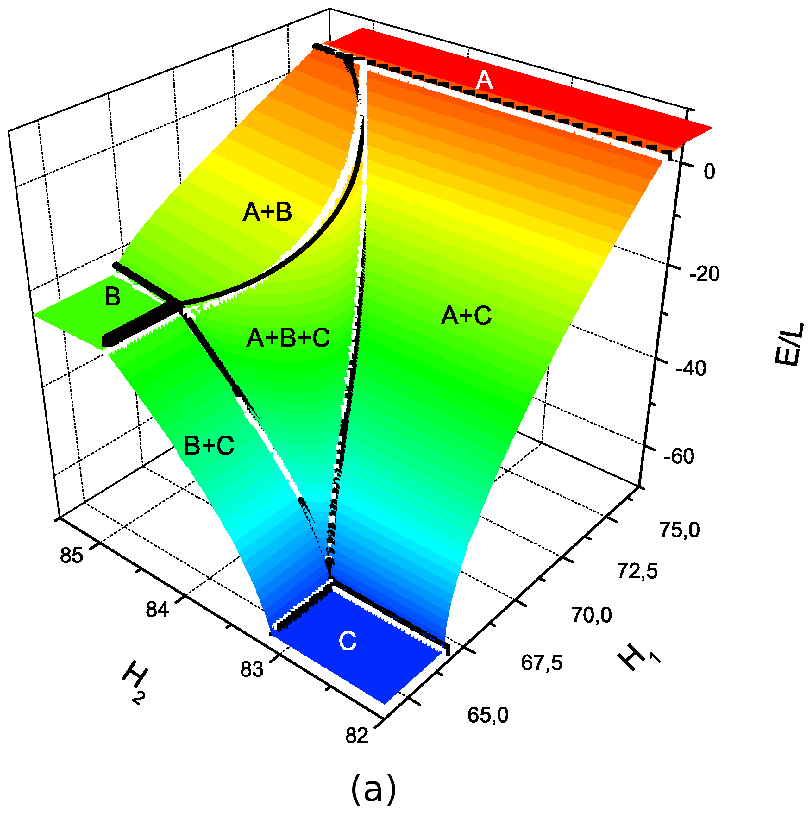}}}
{{\includegraphics [width=0.5\linewidth]{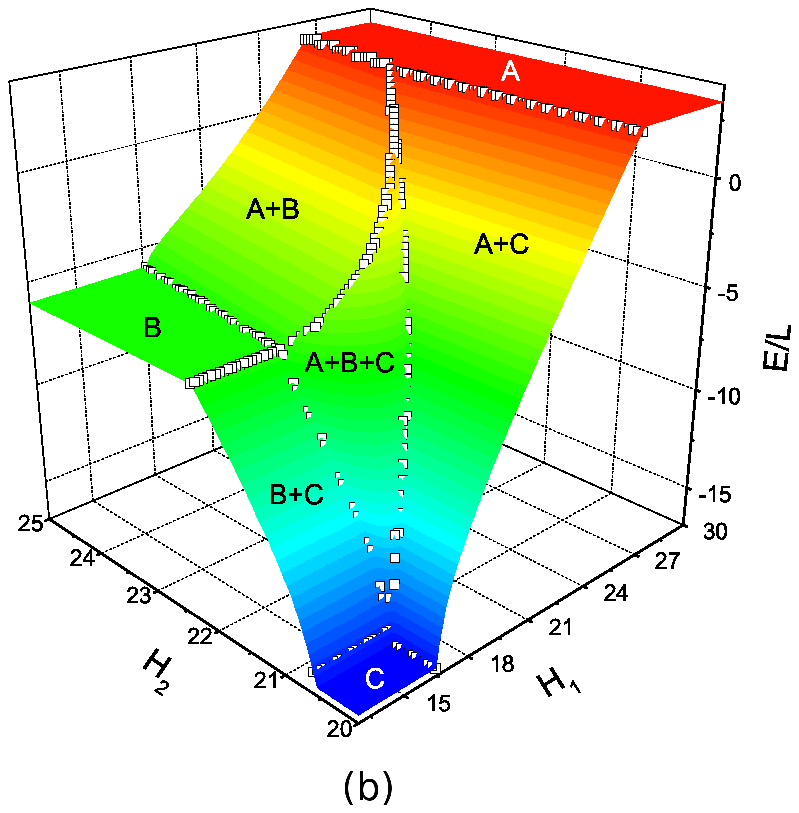}}}\\
{{\includegraphics [width=0.5\linewidth]{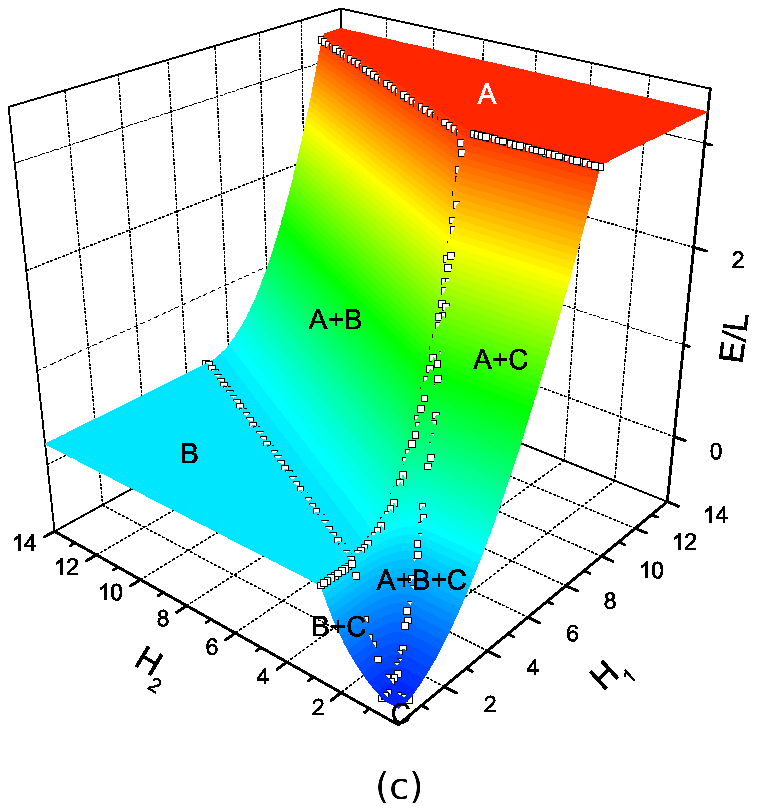}}}
{{\includegraphics [width=0.5\linewidth]{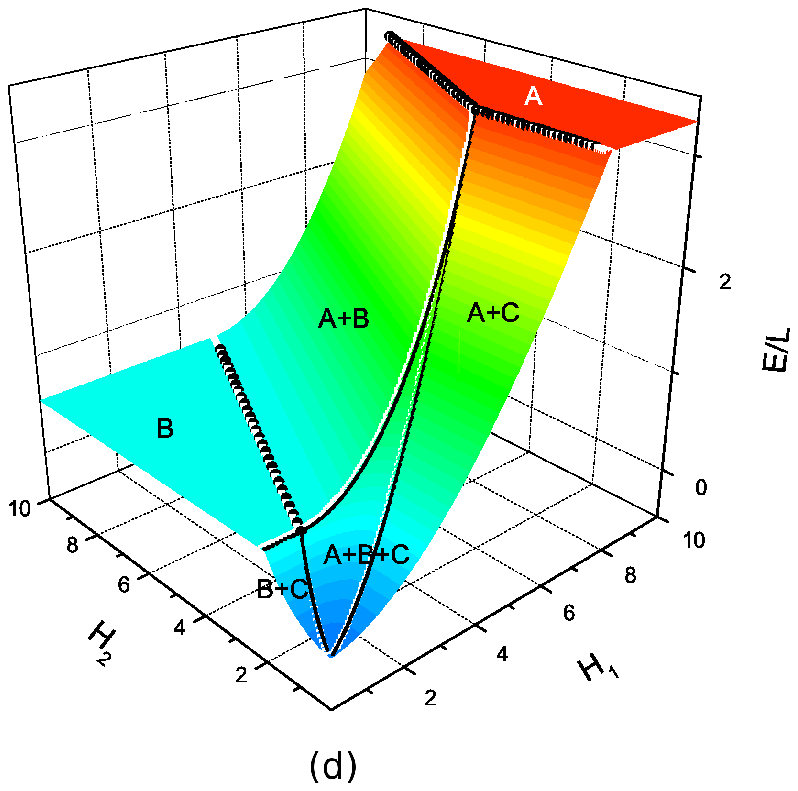}}}
\caption{Ground state energy {\em vs} Zeeman splitting for different coupling 
values (a) strong interaction $|\gamma| = 10$, (b) $|\gamma|=5$, (c) $|\gamma|=1$ and (d) weak interaction $|\gamma|=0.5$. 
The white dots correspond to the numerical solutions of the dressed energy 
equations (\ref{eps}). 
The black lines in (a) and (d) are plotted from the analytical results (\ref{H1-H2}) and (\ref{hweak}), respectively.
A good agreement is found between the analytical results and the 
numerical solutions in the strong and weak regimes. The pure trionic phase $C$, present in the strong coupling 
regime, reduces smoothly as $|\gamma|$ decreases and is suppressed in the weak limit.}
\label{fig:E} 
\end{figure}
Fig. \ref{fig:E}(a) shows the ground state energy versus Zeeman splitting parameters $H_1$ and $H_2$ determined 
from equation (\ref{E}) with the densities $n_1$ and $n_2$ obtained from (\ref{H1-H2}).
There are three pure phases: an unpaired phase $A$, a pairing phase $B$ and a trion phase $C$ and four different mixtures of these states.
For small $H_1$, a transition from a trionic state into a mixture of trions and pairs occurs as $H_2$ exceeds 
the lower critical value $H_2^{c1}$.
When $H_2$ is greater than the upper critical value $H_2^{c2}$, a pure  pairing phase takes place.
Trions and BCS-like pairs coexist when $H_{2}^{c1}<H_2<H_{2}^{c2}$.
These critical fields, derived from equation (\ref{H1-H2}), are given by
$H_{2}^{c1} \approx n^2 \left(\frac{5\gamma^2}{6} - \frac{2\pi^2}{81}(1+\frac{8}{27|\gamma|}-\frac{1}{27\gamma^2})\right)$ 
and $H_{2}^{c2} \approx n^2 \left(\frac{5\gamma^2}{6} + \frac{\pi^2}{8}(1+\frac{20}{27|\gamma|}-\frac{1}{36\gamma^2}) \right)$.
The phase transitions from $B \rightarrow A+B \rightarrow A $ induced by increasing $H_1$ are reminiscent of 
those in the two-component systems \cite{GBLB,Penc}. Basically in this region the highest level is far away 
from the other two levels, so the system reduces to the spin-1/2 fermion case. The mixed phase containing BCS-like pairs 
and unpaired fermions can be called a FFLO phase. We mention that a 
discussion about the pairing nature of 1D many-body 
systems can be found, for instance, in \cite{31,32}.
For small $H_2$, a phase transition from a trionic into a mixture of trions and unpaired fermions occur. 
Using equation (\ref{H1-H2}), we find that the trionic state with zero polarization $n_1/n = 0$ forms the ground 
state when the field $H<H_{1}^{c1}$, 
where $H_{1}^{c1}\approx n^2 \left( \frac{2\gamma^2}{3} - \frac{\pi^2}{81}(1+\frac{4}{9|\gamma|}+
\frac{1}{9\gamma^2})\right)$. When $H_1$ is greater than the upper critical value $H_{1}^{c2} 
\approx n^2 \left(\frac{2\gamma^2}{3} + \pi^2(1-\frac{4}{9|\gamma|}) \right)$, 
all trions are broken and the state becomes a normal Fermi liquid. 

\begin{figure}[t]
{{\includegraphics [width=0.33\linewidth]{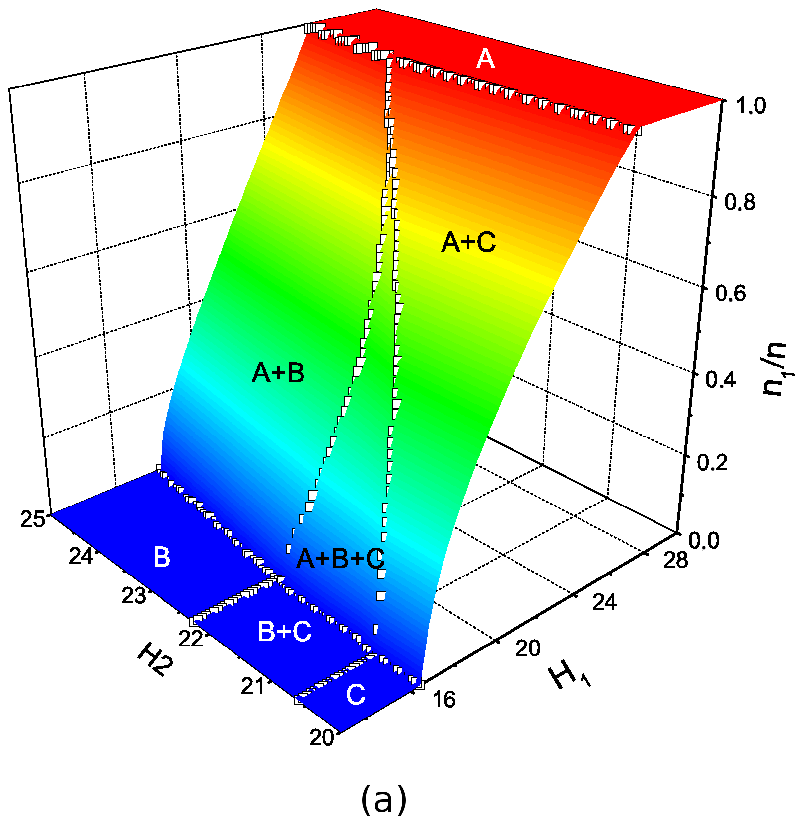}}}
{{\includegraphics [width=0.33\linewidth]{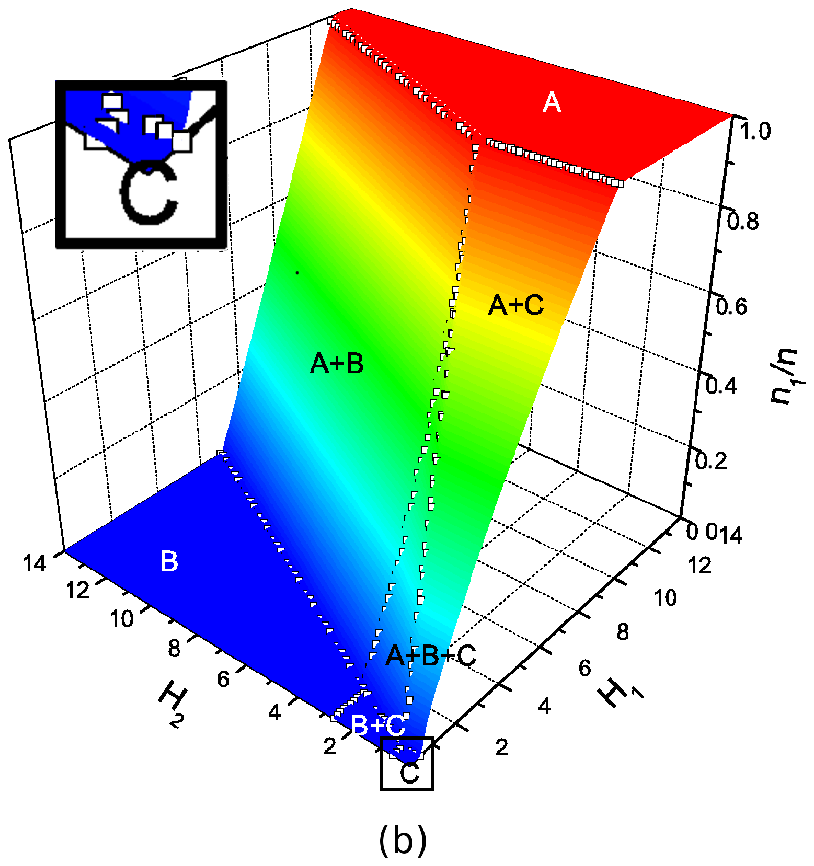}}}
{{\includegraphics [width=0.36\linewidth]{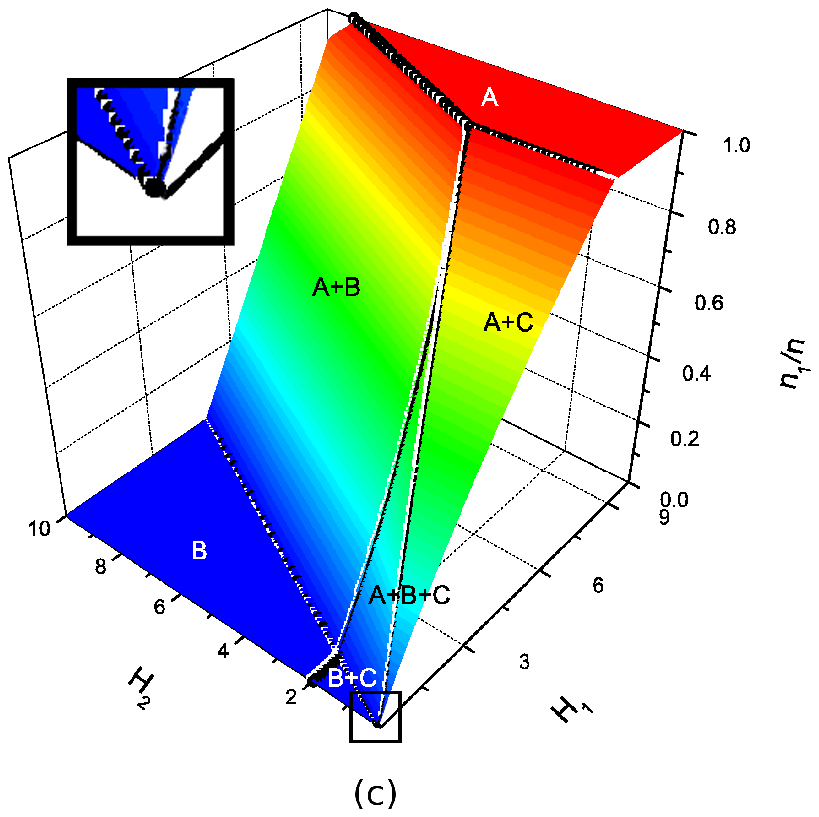}}}
\caption{Phase diagram showing the polarization $n_1/n$ versus the fields $H_1$ and $H_2$ 
for different coupling values (a) $|\gamma| = 5$, (b) $|\gamma| = 1$ and (c) weak interaction $|\gamma| = 0.5$. The white dots 
correspond to the numerical solutions of the dressed energy equations (\ref{eps}). A good agreement is found between the 
analytical results (\ref{hweak}) (black lines) and the numerical solution in the weak regime. At intermediary coupling regimes, 
the pure trionic phase reduces by decreasing the coupling and it is not present in the weak regime. 
A zoom around the origin is presented in Fig. 2(b) (Fig. 2 (c)) to show the presence (absence) of the trionic phase.} 
\label{n1weak}
\end{figure}

\begin{figure}[t]
{{\includegraphics [width=0.33\linewidth]{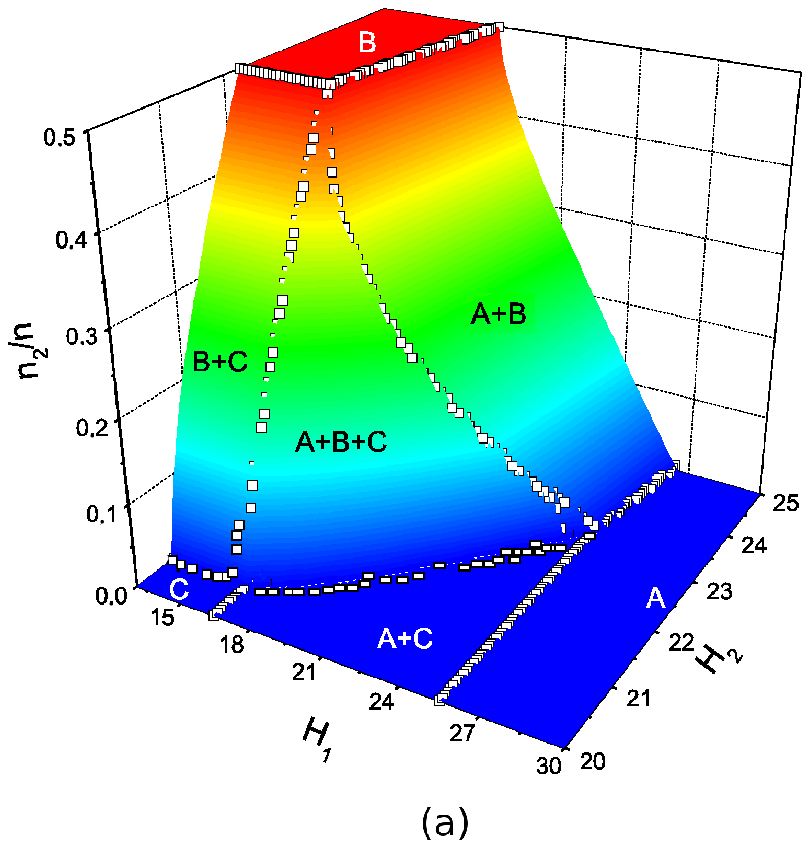}}}
{{\includegraphics [width=0.33\linewidth]{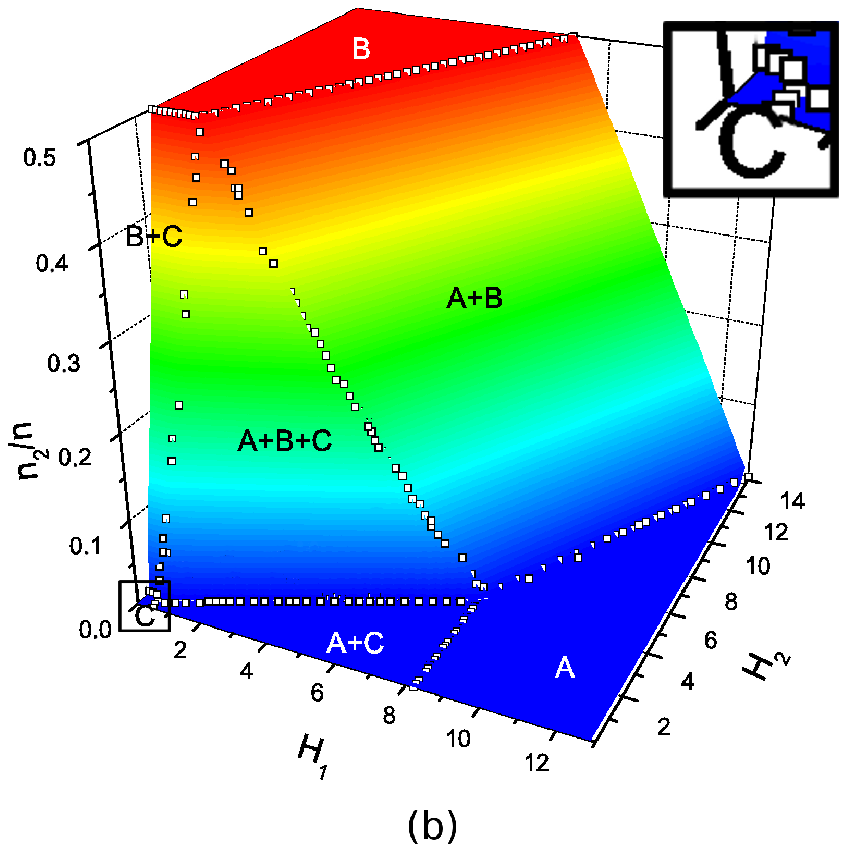}}}
{{\includegraphics [width=0.33\linewidth]{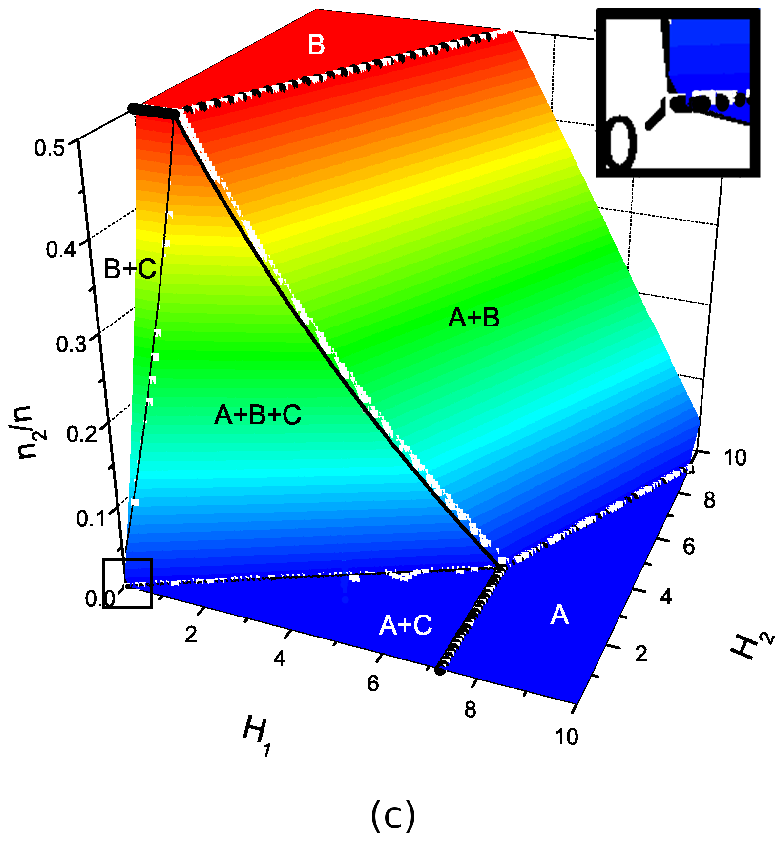}}}
\caption{Phase diagram showing the polarization $n_2/n$ versus the fields $H_1$ and $H_2$ for different 
coupling values (a) $|\gamma| = 5$, (b) $|\gamma| = 1$ and (c) weak interaction $|\gamma| = 0.5$. 
The white dots correspond to the numerical solutions of the dressed energy equations (\ref{eps}). The black 
lines plotted from the analytical results (\ref{hweak}) are in good agreement with 
the numerical results in the weak regime. At intermediary coupling regimes, the pure trionic phase 
reduces by decreasing the coupling and it is not present in the weak regime. A zoom around 
the origin is presented in Fig. 3(b) (Fig. 3(c)) to show the presence (absence) of the trionic phase.} 
\label{n2weak}
\end{figure}

At intermediate coupling regimes, it is not possible to construct the full phase diagrams analytically. 
However, they can be determined by numerically solving the dressed energy equations (\ref{eps}), as 
illustrated in Figs. \ref{fig:E}(b), \ref{fig:E}(c);  \ref{n1weak}(a), \ref{n1weak}(b) 
and \ref{n2weak}(a), \ref{n2weak}(b), for the intermediary values of the coupling $|c|=5$ and $|c|=1$, respectively.
The different phase boundaries modify slightly by varying the inter-component coupling through the whole 
attractive regime. In particular, the pure trionic phase existing in the strong coupling regime reduces 
smoothly by decreasing this coupling until it is completely suppressed. A careful numerical analysis of the 
phase diagrams for $n=1$ and different values of $|c|$ between $|c|=1$ and $|c|=0.5$ indicates that the 
critical coupling value at which the trionic 
phase disappears is around $c_c\approx0.6$. Other mixed phases involving trions, 
specially the phase $(B+C)$ also reduce by decreasing $|c|$.


In the weak coupling regime we obtain the expressions between the fields and the polarizations 
using equations (\ref{Eweak}), (\ref{termod}) and (\ref{constraint}) 

\begin{eqnarray}
H_1 &=& \frac{\pi^2}{3}(2n_1^2 + n_2^2 + 4n_1n_2 +4n_1n_3 + 2n_2n_3)\nonumber\\
&+&\frac{2|c|}{3}(2n_1+n_2). \nonumber\\
H_2 &=& \frac{\pi^2}{3}(n_1^2 + 2n_2^2 + 2n_1n_2 +2n_1n_3 + 4n_2n_3)\nonumber\\
&+&\frac{2|c|}{3}(2n_2+n_1).
\label{hweak}
\end{eqnarray}
These equations together with the energy-field transfer relation $H_1-H_2/2$ determine the full phase diagram and the 
critical fields activated by the Zeeman splitting $H_{1}$ and $H_{2}$.
We observe that the density of trions $n_3$ does no appear independently in eqs.(\ref{hweak}), in contrast 
to the corresponding equations in the strong regime (\ref{H1-H2}).
Fig. \ref{fig:E}(d)  presents the ground state energy versus the fields $H_1$, $H_2$ while 
Figs. \ref{n1weak}(c) and \ref{n2weak}(c) show the polarizations $n_1/n$ and $n_2/n$ in terms of Zeeman splitting, respectively.
Now in the weak coupling regime there are just six different phases in the $H_1$--$H_2$ plane: we observe 
the disappearance of the pure trionic phase C in the presence of the fields, i.e., the trionic state 
is unstable against thermal and spin fluctuations.
This behaviour is in contrast to the strong coupling regime, where the phase $C$ is robust  and trion states populate the 
ground state for a considerable interval of the fields. In addition, the phase where trions and pairs coexist ($B+C$) 
reduces significantly compared to the strong coupling regime. 
Interestingly, in contrast to the weak attractive spin-1/2 fermion system, a pure paired phase can be sustained 
under certain Zeeman splittings. For certain tuning $H_1$ and $H_2$, the two lowest levels are almost degenerate. Therefore, the paired
phase naturally  occurs and is stable. 
The persistence of this phase is relevant for the investigation of phase transition between BCS-like pairs and FFLO states.
All these boundary modifications occur smoothly, as shown by a numerical analysis of the phase diagrams for different 
values of the coupling across all regimes. This indicates that all phase transitions in the vicinities of critical points are second order. 
This conclusion is consistent with previous analytical results \cite{Penc,GBLB}. We also mention that quantum phase 
transitions between different superfluid phases have been discussed in \cite{catelani}.
%
%
%

We perform a similar analysis as in the previous strong case to extract the critical fields.
Since the trionic phase C disappears for non-vanishing fields, less critical fields are found compared to the strong coupling case.
For small $H_1$  a transition from a mixture of trions and pairs into a pure paired phase occurs as $H_2$ exceeds the critical value
$H_{2}^{c} \approx n^2 \left(\frac{2\gamma}{3} + \frac{\pi^2}{6}\right).$ 
The transition from a mixture of trions and unpaired fermions into a normal Fermi liquid phase occurs as $H_1$ exceeds 
the critical value $H_{1}^{c} \approx n^2 \left(\frac{4\gamma}{3} + \frac{2\pi^2}{3}\right).$
The phase transitions $B \rightarrow A+B \rightarrow A $ are reminiscent of those in spin-1/2 fermion systems.
However, in the weak attractive two-component case the pure BCS-like paired phase is suppressed \cite{jingsong} and 
consequently it is not possible to investigate the phase separation between a BCS-like paired phase and a FFLO state, 
in contrast to the three-component case, where this study is still possible in a weak regime.
\section{Conclusion}

We have studied the three-component attractive 1D Fermi gas in external fields through the Bethe ansatz formalism. 
New results for the critical fields and complete zero temperature phase diagrams have been presented for the weak 
coupling regime. Previous work on this model has been extended to derive higher order corrections to these physical 
quantities in the strong regime. We have further confirmed that the system exhibits exotic phases of trions, bound pairs, 
a normal Fermi liquid and mixture of these phases in the strongly attractive limit. 
We have also shown how the different phase boundaries deform by varying the inter-component coupling across 
the whole attractive regime.
In particular, the trionic phase that may occur in the strong coupling regime for certain values of the 
Zeeman splittings reduces smoothly by decreasing the coupling until the weak limit is approached, when 
the trionic phase is suppressed.
Interestingly, in the weak regime, a pure paired phase can be maintained under certain nonlinear Zeeman 
splittings, in contrast to the two-component attractive 1D Fermi gas.
Our high precision of critical phase boundaries pave the way to further
investigate quantum criticality in three-component interacting Fermi gas through the
finite temperature Bethe ansatz.
%
%
%
%
\section{Acknowledgments}

This work has been supported by CNPq (Conselho Nacional de
Desenvolvimento Cientifico e Tecnol\'ogico). The authors would like
to thank X.W. Guan for helpful discussions and comments.
\end{document}